\documentclass[preprint,preprintnumbers,amsmath,amssymb]{revtex4}
\usepackage{subfigure}
\usepackage{amssymb}
\usepackage[dvips]{color}
\usepackage{setspace}
\usepackage{natbib}
\usepackage{url}
\usepackage{bm}
\usepackage{textcase}
\usepackage{color}
\usepackage{graphicx}
\usepackage{amsmath}
\usepackage{ulem}
\usepackage[colorlinks=true,citecolor=blue]{hyperref} 
\usepackage[utf8]{inputenc}
\hypersetup{pdftitle=Theoretical Study of Ternary CoSP Semiconductor: a Candidate for Photovoltaic Applications}
\hypersetup{pdfauthor={A. Houari and F. Benissad}}
\hypersetup{pdfdisplaydoctitle}
\begin{document}
\title{Theoretical Study of Ternary CoSP Semiconductor: a Candidate for Photovoltaic Applications}
\author{Abdesalem Houari}
\email[corresponding author:]{abdeslam.houari@univ-bejaia.dz}
\affiliation{Theoretical Physics Laboratory, 
             Department of Physics, 
             University of Bejaia, 
             Bejaia, Algeria}
\author{Fares Benissad}
\affiliation{Theoretical Physics Laboratory, 
             Department of Physics, 
             University of Bejaia, 
             Bejaia, Algeria}
\date{\today} 
\begin{abstract}
  The electronic structure of pyrite-type cobalt phosphosulfide (CoSP)
  has been studied using density-functional theory.  The calculated
  band structure reveals the non-magnetic semiconducting character of
  the compound. The electronic structure is described through the
  electronic band structure and the densities of states. A band gap of
  1.14 eV has been computed within standard GGA, a value which is
  enhanced using hybrid functional. It separates the upper part of the
  valence band dominated by Co-{\it 3d-t$_{2g}$} states from the lower
  part of the conduction band made exclusively of Co-{\it 3d-e$_g$},
  above of which lie S-{\it 3p} and P-{\it 3p} ones. The obtained
  values are suitable for applications in solar cells, according
  to Shockley-Queisser theory of light to electric conversion
  efficiency. The origin of the larger CoSP band gap, with respect to
  the one of the promising FeS$_2$ compound, is explained and the
  chemical bonding properties are addressed. A comparative picture
  is established where several similarities have been found, suggesting
  that CoSP could be for a great practical interest in photovoltaics.
\end{abstract}
%
%
\maketitle 
\tableofcontents
%
\section{Introduction}
\label {I}
  Pyrite-type iron disulfide (FeS$_2$) is considered as a promising
  material for photovoltaic applications
  \cite{ennaoui93_semsc29_289,wadia09_est43_2072,acevedo14_jacs136_17163}.
  The non toxicity and the abundance on the earth surface, as a
  mineral
  \cite{ennaoui86_jes133_97,birkholz91_prb43_11926,murphy09_ssr64_1},
  make this compound very interesting for practical usage and
  technology.  Another important advantage is its high optical
  absorption \cite {ferrer90_ssc74_913}, which is one of the key
  requirement in solar cell applications. The role of surface states,
  defects and non-stoichiometry with respect to their implications on
  the photovoltaic performance have been recently addressed, by means
  of {\it ab-initio} methods \cite{sun11_prb83_235311,sun11_prb84_035212,eyert98_prb57_6350}.
  Moreover, FeS$_2$ has attracted much interests in other areas. For
  instance, it has been investigated in recent years as a hydrogen
  evolution reaction
  (HER)catalyst\cite{kong13_ees6_3553,faber14_jpcc118_21347}.  It was
  also found that it could be a candidate for thermoelectricity, due
  to it's good thermopower, compared to other materials intended for
  practical use \cite{bither68_ic7_2208,gudelli13_jpcc117_21120}.
  The efficiency and the peformance of a thermoelectric material
  is characterized by its figure of merit (zT) which is proportional
  to the square of the thermopower (S$^2$).

  For photovoltaic applications, however, an important handicap
  for single junction cells applications
  comes from its gap magnitude (experiments give 0.9~eV) which is
  smaller than the optimal value predicted by Shockley-Queisser
  theory \cite{shockley61_jap32_510}.  Recently, tabulated values
  for the maximum efficiency of light to electric power conversion
  have been provided as function of the solar cell band gap
  \cite{ruhle16_se130_139}. The highest one of $\sim 33.16\%$ is
  reached at 1.34~eV of this latter, and conversion efficiencies
  above $30\%$ can be obtained for band gaps between 1~eV and
  1.7~eV. From this, the FeS$_2$ one is 0.44~eV smaller than the
  optimal one. In this context, a lot of efforts have been devoted
  to increase its gap and to improve its other physical
  properties\cite{sun11_prb84_245211}.  A well known scheme for
  tuning electronic properties of semiconductors, such as band gap,
  is alloying via substitution\cite{wei_jap78_3846}.

  In pyrite-type transition metal disulfides TS$_2$ (T=Mn, Fe, Co
  and Ni), it's rather the metallic substitution which is mostly
  considered and commonly explored \cite{houari18_JPCM30_305501}.
  Particularly, a large variety is reported in literature about iron
  pyrite giving rise to Fe$_{1-x}$T$_x$S$_2$ alloys with different
  properties (see
  \cite{houari18_JPCM30_305501,sun11_prb84_245211,khalid15_jmcc46_12068}
  and references therein). On the other side, however, much less
  attention has been given to non-metallic {\it i.e.} sulfur
  substitution.  Nevertheless, alloying with oxygen to form
  FeS$_{2-x}$O$_x$ has been recently reported \cite{wu2016method}.
  By replacing about 10\% of the sulfur atoms, a band gap in a range
  from about 1.2~eV to 1.3~eV has been obtained. In addition to some
  few investigations in electrocatalysis
  \cite{wu17_acsc7_4026,li16_nr10_814}, there is no much reports
  dealing with light elements, such as first and second row
  elements (C, O, Al, Si, P ..etc), replacing sulfur.

  In view of the growing interest to enhance FeS$_2$ properties, it
  appears highly desirable to find new candidates with similar
  interest and suitable properties for photovoltaics and
  thermoelectrics. Starting from the neighboring isostructural
  cobalt disulfide CoS$_2$, a hardly known semiconductor was
  identified in early 1960s by replacing one sulfur atom with one
  phosphorus {\it i.e.} the cobalt mono-phosphosulfide
  Co$_1$S$_1$P$_1$ \cite{hulliger63_nat198_382}.  In recent
  experimental study, Acevedo {\it et al.} \cite
  {acevedo15_nm14_1245} have confirmed that this compound remains in
  the pyrite crystal structure, and could be for a great interest as
  high-performance catalyst for hydrogen production.  According to
  the authors, its crystal structure can be described by Co$^{3+}$
  octahedra and {\it S-P} dumbbells, with a homogeneous distribution
  of P$^{2-}$ and S$^{1-}$ atoms.  Having one electron less than
  CoS$_2$, ternary CoSP is isoelectronic to FeS$_2$, which is
  consistent with the observed semiconducting character. The
  reported lattice constant ({\it a}=5.422\AA)
  \cite{hulliger63_nat198_382} is almost equal to the FeS$_2$ one
  ({\it a}=5.415\AA), while being smaller than the CoS$_2$
  corresponding constant ({\it a}=5.538\AA).  Except these mentioned
  structural data, we did not find much works in literature on
  this cobalt phosphosulfide compound. Its physical properties
  (electronic, optical ..etc ) remain unexplored both experimentally
  as well as theoretically.

  To study correctly the optical properties (usually more difficult
  within standard DFT and need to go beyond one particle
  approximation), a complete and clear picture of the electronic
  structure is necessary beforehand.  That is, we restrict our present
  study to the electronic properties of CoSP compound, through the
  computation of band structure and densities of states. Further
  insights are given by comparing the results with well established
  ones of FeS$_2$.
  
 The paper is organized as follows: In Sec.~\ref{paw}, we describe the
 theoretical method and the computational details.  In
 Sec.~\ref{results} we present our results and a summary is finally
 given in Sec.~\ref{conclusion}
     
\section{Computational Method}
\label{paw}
  The {\it ab-initio} calculations are carried out in the framework of
  density-functional theory
  (DFT)\cite{hohenberg64_pr136_B864,kohn65_pr140_1133}. The electronic
  properties (band structure and densities of states), have been
  obtained with the Projector Augmented Wave (PAW)
  method\cite{bloechl94_prb50_17953} as implemented originally in the
  CP-PAW code. The PAW method is an {\it all-electron} electronic
  structure method, where the full wave functions including their
  nodal structure are properly defined. The PAW formalism
  combine and generalize ideas of both pseudopotentials
  (USPP)\cite{vanderbilt90_prb41_7892} and the linearized
  augmented-plane-wave (LAPW) method \cite{andersen75_prb12_3060}.
  The method recovers the missing link between both approachs. Compared
  to the former, PAW is more rigorous, and can be
  made exact by converging series expansions. A thorough description
  of the method can be found in the original paper by P. E. Bl\"ochl
  \cite{bloechl94_prb50_17953}.

  The exchange-correlation effects are accounted within the
  generalized gradient approximation (GGA) of Perdew-Burke-Ernzerhof
  \cite{perdew96_prl77_3865}.  Moreover, we also performed
  calculations using the hybrid PBE0r functional, which replaces a
  fraction of the exchange term of the PBE functional with the exact
  Fock-term\cite{becke93_jcp98_1372,perdew96_jcp105_9982}.  Unlike in
  PBE0\cite{adamo99_jcp110_6158}, it uses the idea of range separated
  (screened) hybrid functional\cite{heyd03_jcp118_8207}, where the
  slowly decaying long-range part of the Fock exchange interaction is
  replaced by the corresponding part of the PBE counterpart. The
  PBE0r functional has been developed by Bl\"ochl {\it et al.}
  \cite{bloechl11_prb84_205101,bloechl13_prb88_205139}, where the
  range separation correction is restricted to the onsite interactions
  in a local orbital basis. The scaling of the onsite terms
  combined with the truncation of off-site terms approximates the
  screening of the interaction in the exchange term in the spirit of
  the random phase approximation. The double counting term or the
  removal of the GGA exchange is done consistent with choice of the
  screened U-tensor used in the Fock term.
  \cite{bloechl11_prb84_205101}. The mixing parameter for the exchange
  can be chosen independent for each atom. So far, the best agreement
  with spectral properties has been obtained with a mixing factor
  close to $\frac{1}{8}$, and PBE0r functional has shown to be
  accurate for transition-metal oxides with a partially filled d-shell
  \cite{sotoudeh17_prb95_235150}.

  CP-PAW employs the framework of Car-Parrinello {\it ab-initio}
  molecular dynamics (AIMD)\cite{car85_prl55_2471} for the
  optimization of wave functions and atomic structure. It is based on
  a fictitious Lagrangian from which a set of Euler-Lagrange type
  equations of motion are derived for the electronic wavefunctions as
  well as for the atomic positions.  In the Car-Parrinello method, the
  electronic wavefunctions and atomic positions are treated on an
  equal footing, and the groundstate is simulated by applying
  friction. 

  For the augmentation, we used a $s^1p^1d^1$ set of projector
  functions for all atoms, where the superscripts denotes the number
  of projector functions angular momentum channel. The convergence of
  the total energy minimization is reached when the difference between
  two successive iterations is less than 10$^{-5}$Hartree (also known as
  the self-consistency convergence criterion). On the other
  hand, the convergence of the total energy versus two important parameters,
  which are the plane wave cutoff (wavefunction and density) and the number of
  {\it k}-points in the Brillouin-zone, has been also obtained carefully. 
  For the former, several increasing cutoff values (30, 40, 60 up to 70~Ry)
  have been considered. With values higher than 40~Ry, the variation in the
  computed total energy is less 10$^{-4}$ Hartree which is always considered
  as a very good accuracy in the field of DFT calculations. The Brillouin-zone
  integration has been performed with the linear tetrahedron
  method\cite{jepsen71_ssc9_1763, lehmann72_pssb54_469} and the
  so-called Bl\"ochl corrections\cite{bloechl94_prb49_16223}.  As for
  plane wave cutoff, in order to ensure the convergence of the total
  energy versus the {\it k}-points sampling, the calculations have been carried
  out with increasing grids (from $4\times4\times4$ up to
  $7\times7\times7$).  With a $6\times6\times6$ mesh, leading to 112
  {\it k}-points in the irreducible Brillouin-zone, an accuracy of
  10$^{-4}$ Hartree has been reached. All structural parameters
  (lattice constant and atomic positions) have been optimized.

\section{Results and Discussions}
\label{results}

\begin{table}[h]
  \caption{Calculated lattice constant ({\it a}) and bond lengths 
    (in \AA) within GGA for $\rm {CoSP}$. For sake of comparison, the corresponding
    data for iron and cobalt disulfides  ($\rm {FeS_2}$ and $\rm {CoS_2}$) are shown. 
    Experimental data\cite{fujimori96_prb54_16329,acevedo15_nm14_1245} are given in 
    parentheses.}
  \label{tbl:table1}
\begin{ruledtabular}
\begin{tabular}{lll}
                           & $ a$[\AA]     &  Bond lengths ([\AA]) \\
\hline
$ {\rm CoSP} $             & 5.411 (5.422)  & d$_{\rm Co-S}$=2.278   \\
                           &                & d$_{\rm Co-P}$=2.266   \\
                           &                & d$_{\rm S-P}$=2.106    \\
\hline
$\rm FeS_2$                & 5.405 (5.416)  &  d$_{\rm Fe-S}$=2.263 \\
                           &                &  d$_{\rm S-S}$=2.160  \\
\hline
$\rm CoS_2$                & 5.489 (5.538)  &  d$_{\rm Co-S}$=2.310 \\
                           &                &  d$_{\rm S-S}$=2.101   \\
\end{tabular}
\end{ruledtabular}
\end{table}

  We address here the structural and electronic properties of the
  cobalt phosphosulfide $\rm {CoSP}$. In the following, we give an
  exhaustive discussion of the electronic band structure and
  densities of states (DOS). Simultaneously, differences and
  similarities with the well established electronic structure of $
  {\rm FeS_2}$ are highlighted.

  The calculated lattice constant and bond lengths (Co-S, Co-P and
  S-P) within GGA are given in Table \ref{tbl:table1}, in addition
  to the calculated corresponding data of $\rm {FeS_2}$ and $\rm
  {CoS_2}$. The latter, as well as the experimental data shown
  between parentheses, are included for sake of comparison. The
  calculated values are in very good agreement with experimental
  ones. The results obtained with hybrid functional are very close,
  (only 0.1\% larger) and consequently have been omitted in the
  table.  As it can be noticed, the Co-S(P) bond lengths lie between
  the corresponding ones of the pure compounds, being much closer to
  Fe-S than to Co-S.  This is consistent with the lattice constant
  values of the three compounds.  On the other hand, however, the
  non-metallic S-P dumbbell distance in $\rm {CoSP}$ is almost equal
  to the S-S one in $\rm {CoS_2}$. It is relatively shorter than the
  corresponding one in $\rm {FeS_2}$, and this will have very
  important effects, especially on the gap property.    

\begin{figure}[h]
\includegraphics[width=10cm, height=8cm]{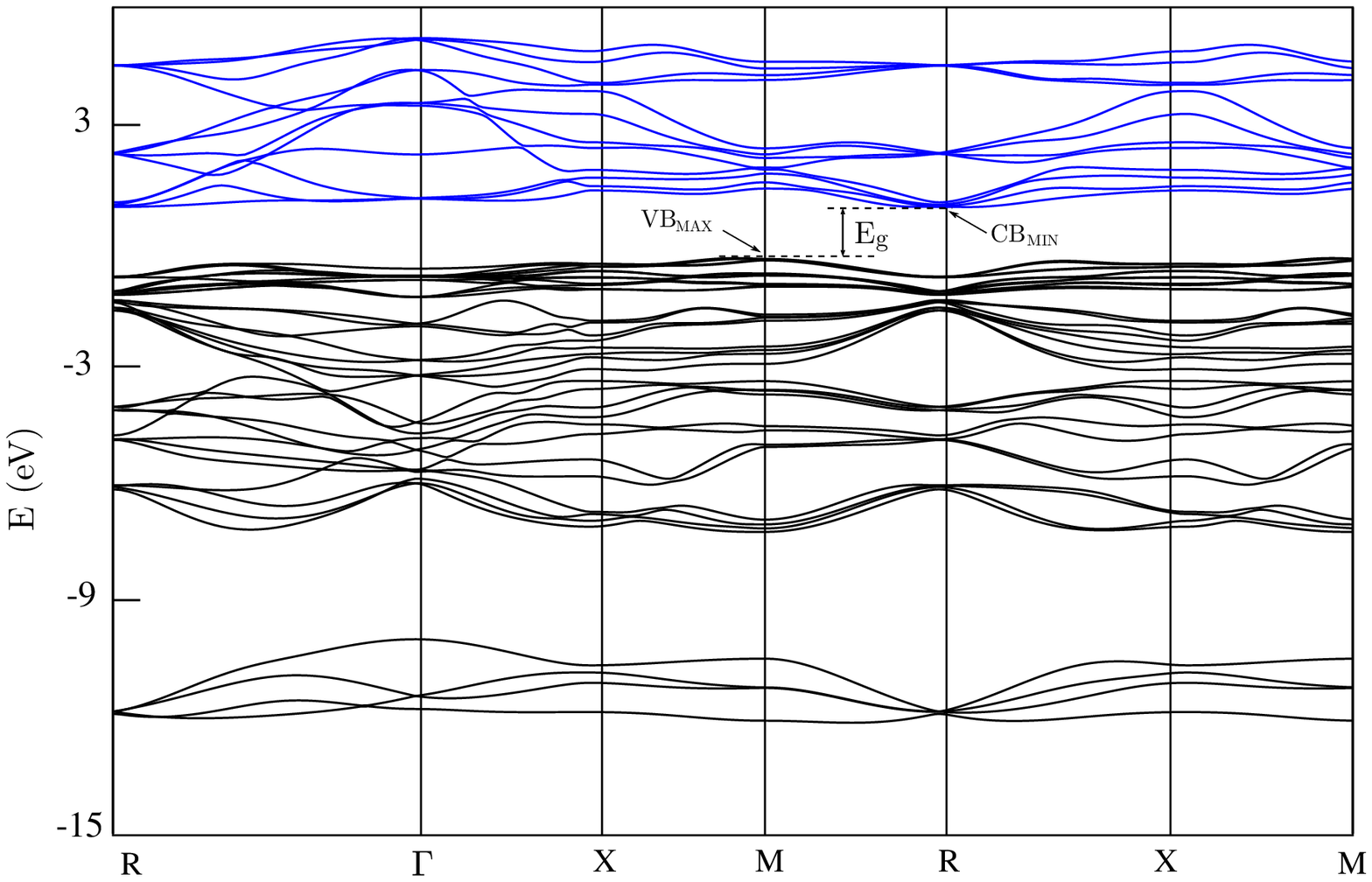}
\caption{(Color online) Electronic band structure (eV) of
  $\rm {CoSP}$ along selected high symmetry lines of the 1$\rm ^{st}$
  Brillouin zone of the simple cubic lattice.
  The filled states (core and valence bands) are drawn in black
  color, whereas the empty states (conduction bands) in blue.
  The valence band maximum (VB$\rm _{MAX}$) and conduction band
  minimum (CB$\rm _{MIN}$) are separated by a band gap E$\rm _g$.}
\label{fig:gga-bands}
\end{figure}

  It has been shown that a very small change of the S-S bond length in
  $\rm {FeS_2}$ could influence drastically the band gap magnitude
  (and even its nature) \cite{eyert98_prb57_6350}. Only small
  shortening could lead to a substantial increase of the band gap. In
  view of the structural results (see Table \ref{tbl:table1}), we
  expect an enhanced band gap in $\rm {CoSP}$ compared to $\rm
  {FeS_2}$. As a matter of fact, this is confirmed in a first step
  within GGA framework. The plots of electronic band structure along
  the first Brillouin zone high-symmetry points of the simple cubic
  unit cell is illustrated in Fig.\ref{fig:gga-bands}. The filled
  (core and valence) bands are shown in black color and the empty
  conduction bands are shown in blue color. The maximum of the former
  VB$_{\rm MAX}$ is separated by a gap $\rm E_g \sim 1.14$~eV from the
  minimum of the latter CB$_{\rm MIN}$. The individual contribution of
  the atomic species cobalt, sulfur and phosphorus to the bands
  occupation will appear in the next paragraph through the DOS plots.

  From Fig.\ref{fig:gga-bands} we notice that the $\rm CoSP$ band gap
  is an indirect one. This latter feature is common to $\rm FeS_2$
  compound, as well as the whole trend of the bands distribution which
  seems very similar (see references \cite{banjara18_aa8_025212,
  houari18_JPCM30_305501}).  These similarities
  could be attributed to the fact that both compounds are
  isostructural, but more important are isoelectronic i,e. they have
  the same valence configuration.  The extra electron brought by cobalt with respect
  to iron, is lost when one sulfur is replaced by phosphorus. Finally,
  we mention that taking in account the spin polarization results in a
  zero magnetic moment of the unit cell, as well as for individual
  atomic species, leading to non magnetic semiconductor.

  The properties of $\rm FeS_2$ are well established both
  experimentally and theoretically. We just recall that it's a
  semiconductor, with an indirect band gap and most of the
  experimental measurements give $\sim$ 0.9~eV
  \cite{ennaoui93_semsc29_289,schlegel76_jpc9_3363,li74_prl3_470} (a
  good account of the computed and measured $\rm FeS_2$ gap is
  reported in reference \cite{banjara18_aa8_025212}). In a previous
  work on $\rm Fe(Mn,Ni)S_2$ alloys \cite{houari18_JPCM30_305501},
  our GGA calculations gave $\rm E_g\sim 0.35$~eV, much smaller than
  experimental value. The filled Fe-t$_{2g}$ states form a narrow
  sub-bands in the top of the valence band, whereas Fe-e$_g$ ones in
  the bottom of the conduction band are empty. An important detail
  to mention, however, is that the lowest edge of the conduction
  band has an S-{\it p} character.

  The main difference in $\rm CoSP$ concerns this latter (CB$_{\rm
    MIN}$ in Fig.\ref{fig:gga-bands}) which is made of Co-{\it e$_g$}
  type making the gap exclusively between cobalt metallic
  states. As pointed out above, the short bond length of S-P dumbbell
  (compared to S-S one in $\rm FeS_2$), shifts up the {\it p}-states
  of sulfur and phosphorus in the conduction band to higher energies
  above the Co-e$_g$ ones, and consequently leads to a wider band
  gap. This mechanism proposed first on the basis of DFT calculations,
  has been exploited in recent experiments where an improved $\rm
  FeS_2$ gap was achieved through oxygen substitution (oxygen is
  isoelectronic of sulfur)\cite{hu12_jacs134_13216}.

  This underestimation of band gaps in solids is a well known
  failure of the Kohn-Sham DFT framework
  \cite{perdew17_pnas114_2801}. One way to cure this discrepancy is
  the use of hybrid functional. With a fraction of 10$\%$ of (exact)
  Fock exchange, we have reproduced the correct value for $\rm
  FeS_2$. Following the same procedure here i,e. assuming the same
  weight of the Fock term, a gap of $\rm E_g \sim 1.65$~eV has been
  calculated. Therefore, we expect the experimental gap to be around
  the latter value, with the GGA's value being a lower limit ($\sim
  1.14$~eV).  Regarding this gap value property in particular, the
  compound could attract a great interest for photovoltaic
  single-junction in solar-cell energy production. As pointed out in
  the introduction, according to Shockley-Queisser theory
  \cite{shockley61_jap32_510,ruhle16_se130_139}, the range between
  1.0~eV and 1.7~eV leads to the closest values of the threshold
  efficiency (obtainted for $\rm E_g \sim1.4$~eV), and thus our $\rm CoSP$
  predicted ones (either with GGA or hybrid functional) lies in this
  interval.

\begin{figure}[h]
\includegraphics[width=10cm, height=8.5cm]{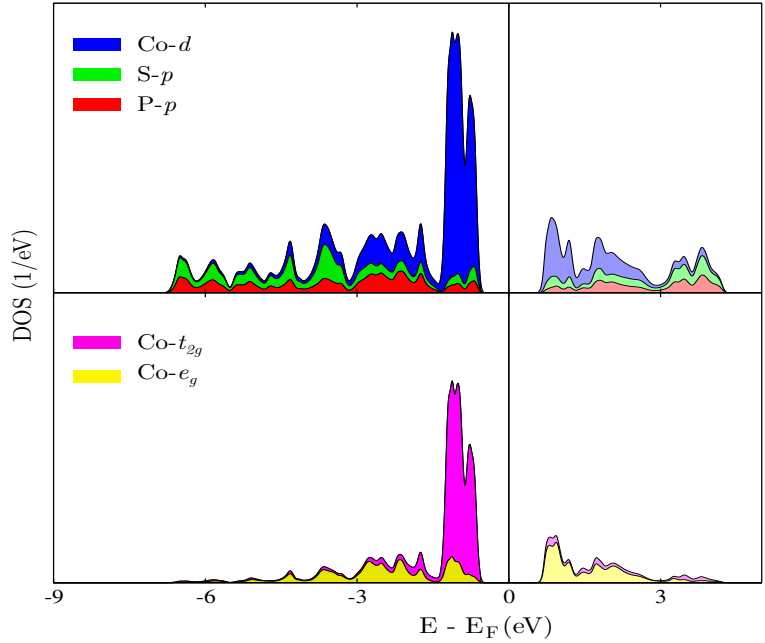}
\caption{(Color online) Density of states (DOS) of  $\rm {CoSP}$
  , where the energy axis origin is set to the Fermi
  level $\rm E_F$. The upper panel: atomic site projected DOS are drawn
  with distinct color as follow: Co-{\it d} , S-{\it p} and P-{\it p}
  states are represented by blue, green and red colors respectively. 
  The lower panel: Co-{\it d} states splitted into $t_{2g}$ (magenta)
  and $e_g$ (yellow) contributions. To illustrate well the atomic
  species contributions individually, each density is drawn in
  the top of another.}
\label{fig:gga-dos}
\end{figure}

  The electronic densities of states (DOS) are shown in
  Fig.\ref{fig:gga-dos}.  Since phosphorus and sulfur are
  neighboring atomic species (i.e, P has one electron less than S ),
  their states lie in the same energy window. The bottom of the
  valence band constituted mainly by {\it s}-states of P and S are
  not shown here (the corresponding bands are, however, visible
  between -9 and -15 eV in Fig.\ref{fig:gga-bands}). As shown in the
  upper panel, the non-metallic contribution (i,e. S and P states)
  dominate from -7 eV to -4 eV, whereas a large part of cobalt
  states are concentrated as one block in narrow energy window
  between -2~eV and-0.8~eV. In addition, the metallic states still
  contribute with a large part, with a small mixture of S and P ones
  in the middle region from -3~eV to -2~eV.  In the lower panel of
  Fig.\ref{fig:gga-dos}, the crystal field splitting of cobalt
  {\it d}-states to its t$_{2g}$ and e$_g$ contributions is illustrated ,
  where a quite clear separation between them is noticeable. The
  former is almost exclusively located in the narrow energy window
  of the top of the valence band with a small admixture of e$_g$
  states.  These latter are in one hand spread in the middle between
  -5~eV and -2~eV, and on the other hand dominate the bottom of the
  empty conduction bands around +2~eV.  Thus, the cobalt t$_{2g}$
  sub-bands are fully occupied, whereas e$_g$ ones are only half
  occupied.

\begin{figure}[h]
\includegraphics[width=10cm, height=7cm]{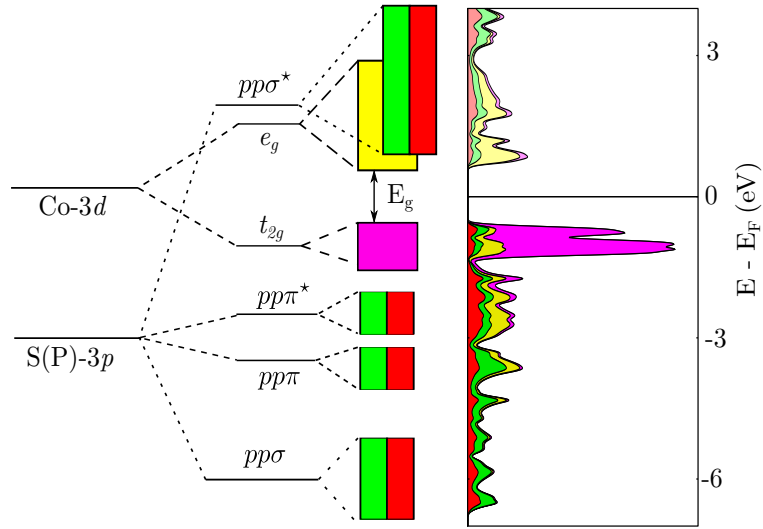}
\caption{(Color online) Qualitative illustration of the chemical
  bonding behavior in $\rm {CoSP}$, through the density of states
  (DOS) plots. A sketch in the form of a molecular
  orbitals diagram showing the main bonding characters of the system.}
\label{fig:mol_orb}
\end{figure}

  The behavior of the chemical bonding in $\rm CoSP$ compound is
  described in Fig.\ref{fig:mol_orb} with the help of a sketch in
  the form of a molecular orbitals diagram. It has to be mentioned that
  this is just a qualitative illustration of the average trend and
  not a quantitative study of the chemical bonding of the system.
  It can be decomposed into three main different regions. In the
  lower part, from -7~eV to -4 ~eV, a clear hybridization between
  {\it p$_z$}-type states of sulfur and phosphorus is involved,
  leading to a strong S(P)-{\it pp$\sigma$} bond. In the middle of
  the valence band (from -4~eV to -1~eV), we find the weaker
  S(P)-{\it pp$\pi$} bond and its counter part antibonding S(P)-{\it
    pp$\pi^{\star}$} one.  This latter is mainly mixed with a small
  fraction of bonding Co-e$_g$ states.  An important non-bonding
  block of Co-t$_{2g}$ orbitals dominate the top of the valence band
  from -1.5~eV until the gap. Finally, an important antibonding
  Co-e$_g$ states occupy the lower part of the conduction bands
  up to +3~eV. They coexist with the antibonding S(P)-{\it
    pp$\sigma^{\star}$} which are spread to higher energies and
  dominate the region above +3~eV.

  Finally, we mention that except the improved value of the gap, the
  behavior and the features of the band structure, the densities of
  states and the chemical bonding obtained with the hybrid functional
  are almost identical to those obtained with GGA. Therefore, both
  frameworks give the same description of the compound and so the same
  discussion holds.

  Before summarizing, we want to point out an important point related to
  the optical properties, already mentioned in $\rm FeS_{2-x}O_x$
  study \cite{hu12_jacs134_13216}. In this latter, the optical
  absorption comes mainly from transitions between metallic states
  (from t$_{2g}$ to e$_g$ states).  It's found that the insertion of
  oxygen replacing sulfur, reducing the local symmetry of Fe-e$_g$
  bands, improves the whole absorption. Such situation is almost
  identical here with $\rm CoSP$, and consequently we expect similar
  interesting optical properties (high optical absorption
  and suitable band gap for efficient light-electricity conversion).
  A study dedicated to these latter
  is in progress. 

\section{Summary and Conclusion}
\label{conclusion}
  To conclude, we explored using first-principles calculations, the
  electronic structure and chemical bonding of $\rm CoSP$
  semiconductor crystallizing in pyrite type crystal structure.

  Being isostructural and isoelectronic to $\rm FeS_2$, interesting
  optical properties like those of this latter are expected. $\rm CoSP$
  compound has been found to have a larger band gap.  It is estimated
  to $\sim$1.14~eV, within GGA and a value of $\sim$1.65~eV is
  obtained using hybrid functional. These values are highly suitable
  for practical usage in solar cell as predicted by Shockley-Queisser
  theory, since they belong to the highest range of light to electric
  conversion efficiencies. Therefore, $\rm CoSP$ could be considered
  as a promising compound for photovoltaic applications, and experimental
  data are called for to confirm our predictions.

  The whole trend of the electronic structure is similar to $\rm
  FeS_2$, where the effect of the crystal field on the cobalt 3{\it
    d}-states is clearly visible. The important hybridization of the
  {\it p$_z$}-states within S-P dumbbell gives rise to a strong
  $\sigma$-bond in the bottom of the valence band, largely splitted
  from its counter part $\sigma^{\star}$-antibond in the conduction
  band. Like in iron pyrite, the cobalt t$_{2g}$ states form a
  narrow non-bonding block dominating the top of the valence band.

  The most important and noticeable feature in $\rm CoSP$ concern
  the lowest edge of the conduction band which is of Co-e$_g$
  character and the larger value of the gap.  This feature is
  attributed to the S-P bond length which is found to play a crucial
  role. Its shorter magnitude (compared to S-S one $\rm FeS_2$)
  shifts up the S(P)-{\it p$_z$} states to higher energy. This
  results in a wider band gap which now separate exclusively
  metallic states i.e, Co-t$_{2g}$ states (VB$_{\rm MAX}$) from the
  Co-e$_g$ ones (CB$_{\rm MIN}$).

\bigskip
\begin{acknowledgements}
  A.H. gratefully acknowledges Prof. Dr. Peter E. Bl\"ochl from
  Clausthal University of Technology (Germany) for several stays
  spent in his group and for fruitful scientific discussions. 
  In particular for having provided us the detailed description
  of their exchange-correlations functional PBE0r of section II
  (Computational Method).
\end{acknowledgements}


\end{document}